\documentclass[12pt]{article}
\usepackage[dvips]{epsfig}
\usepackage{cite}
\newcommand{\ep}{\varepsilon}
\newcommand{\Li}[2]{{\mbox{Li}}_{#1}\left(#2\right)}
\newcommand{\tfrac}[2]{{\textstyle{\frac{#1}{#2}}}}

\newcommand{\Snp}[2]{{\mbox{S}}_{#1\!}\left(#2\right)}

\begin{document}
\renewcommand{\thefootnote}{\fnsymbol{footnote}}

\begin{flushright}
 {BU-HEPP-06-01} \\[3mm]
 {hep-th/0602028} \\[3mm]
 {January 2006}
\end{flushright}
 \vspace*{2.0cm}

 \begin{center}
 {\Large \bf Gauss hypergeometric function: \\ 
reduction, $\ep$-expansion for integer/half-integer parameters and Feynman diagrams}
 \end{center}
\vspace{3mm}
\begin{center}
M.~Yu.~Kalmykov$^{a,b,}$
\footnote{
Supported by 
NATO Grant PST.CLG.980342
and DOE grant DE-FG02-05ER41399
and in part by RFBR grant \# 05-02-17645
and the Heisenberg-Landau Programme}
\\
 \vspace{1cm}
$^{a}${\em
Department of Physics, \\
Baylor University, \\
One Bear Place, Box 97316 \\
Waco, TX 76798-7316}
\\
\vspace{.3cm}
$^{b}${\em
Bogoliubov Laboratory of Theoretical Physics, \\
Joint Institute for Nuclear Research, \\
$141980$ Dubna (Moscow Region), Russia}
\\
\end{center}
  
 \hspace{3in}
 \begin{abstract}
The Gauss hypergeometric functions $_2F_1$ with arbitrary values
of parameters are reduced to two functions with fixed values
of parameters, which differ from the original ones by integers.
It is shown that in the case of integer and/or half-integer values
of parameters there are only three types of algebraically independent
Gauss hypergeometric functions.
The $\ep$-expansion of functions of one of this type 
(type {\bf F} in our classification) demands the introduction of 
new functions related to generalizations of elliptic functions. 
For the five other types of functions the higher-order $\ep$-expansion 
up to functions of {\it weight} {\bf 4} are constructed.
The result of the expansion is expressible in terms of Nielsen polylogarithms only. 
The reductions and $\ep-$expansion of $q$--loop off-shell propagator diagrams with one massive 
line and $q$ massless lines and $q$--loop bubble with two-massive lines and $q-1$ massless lines
are considered. 
\end{abstract}

\vspace{3mm}

%\noindent
%PACS: 11.15.Bt, 02.30.Gp, 02.30.Lt, 12.15.Lk
%%%% extra: 12.38.Bx 
%\\
%Keywords: Feynman diagrams, 
%          Hypergeometric functions, 
%          Higher order $\ep$-expansion,
          
\thispagestyle{empty} 
\newpage
%=====================================================================
\renewcommand{\thefootnote}{\arabic{footnote}}
\setcounter{footnote}{0}

%%%%%%%%%%%%%%%%%%%%%%%%%%%%%%%%%%%%%%%%%%%%%%%%%%%%%%%%%%%%%%%%%%%%%%%%%
\section{Introduction}
\setcounter{equation}{0}
%%%%%%%%%%%%%%%%%%%%%%%%%%%%%%%%%%%%%%%%%%%%%%%%%%%%%%%%%%%%%%%%%%%%%%%
The construction of higher order 
$\ep$-expansions of hypergeometric functions has been intensively discussed in the literature
in the context of the calculation of Feynman diagrams. 
At the present moment, several algorithms for the Laurent expansion of different types of 
hypergeometric functions with respect to small parameter (in the rest of the paper, 
we will call such expansion as $\ep$-expansion) are proposed.
They are mainly related to integer values of the parameters and/or special values of 
the argument. The present paper is concerned with the Gauss hypergeometric function 
\begin{equation}
~_{2}F_1\left(\begin{array}{c|}
A+a\ep, B+b\ep \\
C+c\ep \end{array} ~z \right) 
 = \sum_{j=0}^\infty \frac{(A+a\ep)_j (B+b\ep)_j}{(C+c\ep)_j} \frac{z^j}{j!} \; , 
\label{gauss}
\end{equation}
where
$(\alpha)_j \equiv \Gamma(\alpha+j)/\Gamma(\alpha)$ is the Pochhammer symbol, 
all parameters are real numbers and $\ep$ is a small parameter.
Within dimensional regularization \cite{dimreg}, the parameter $\ep$ is related with 
deviation of d-dimensional space-time from its integer value, $d=m-2\ep$.
Using the well-known representation 
for the Taylor expansion of the Gamma-function for an integer positive number, $m > 1$
\begin{eqnarray}
&&  \hspace{-7mm}
\frac{(m+a\ep)_j}{(m)_j} =
\exp\left\{ -\sum_{k=1}^{\infty} \frac{(-a\ep)^k}{k}
\left[ S_k(m \!+\! j \!-\! 1) \!-\! S_k(m \!-\! 1) \right] \right\} \; ,
\nonumber 
\end{eqnarray}
or half-integer positive values, 
$$
\left(m + \frac{1}{2} + a \ep \right)_{j}=
\frac{\left( 2 m + 1 + 2 a \ep \right)_{2 j} }{4^j \left( m + 1 + a \ep \right)_{j}} \; ,
$$
where 
$S_k(j) = \sum_{l=1}^j l^{-k}$ 
is the harmonic sum~\footnote{
The harmonic sums are related with 
function $\psi(z)=\frac{\rm d}{{\rm d}z}\ln\Gamma(z)$ 
and its derivatives by means of the relation
\[
\psi^{(k-1)}(j) = (-1)^k (k-1)! \left[\zeta_k - S_k(j-1) \right],
  \qquad  k>1,
\]
where $\psi^{(k)}(z)$ is the $k$-th derivative of the $\psi$-function.
In particular, for $k=1$ we have $\psi(j) = S_1(j-1)-\gamma_E$,
and $\gamma_E$ is Euler's constant.
},  
%%%%%%%%%%%%%%%%%%%%%%%%%%%%%%%%%%%%%%%%%%%%%%%%%%%%%%%%%%%%%%%%%%%%%%%%%%%%%%%%%%%%%%%%%%%%
the original hypergeometric function (\ref{gauss}) with integer and/or half-integer values of 
parameters $\{A,B,C\}$ can be written as \cite{MKL04}
\begin{eqnarray}
&& \hspace*{-7mm}
{}_{P\!+\!1}F_P \left(\begin{array}{c|} 
\{ m_i\!+\!a_i\ep \}^{J}, \{ p_j\!+\! \tfrac{1}{2} \!+\! d_j \ep \}^{P\!+\!1\!-\!J} \\ 
\{ n_i\!+\!b_i\ep \}^K,  \{ l_j\!+\! \tfrac{1}{2} \!+\! c_j \ep \}^{P\!-\!K}  
\end{array} ~z \right) = 
\nonumber \\ && \hspace*{-3mm}
\sum_{j=1}^\infty \frac{z^j}{j!} \frac{1}{4^{j(K\!-\!J\!+\!1)}} 
\frac{\Pi_{i=1}^J (m_i)_j}{\Pi_{l=1}^K (n_l)_j}
\Pi_{r=1}^{P\!+\!1\!-\!J} \frac{(2 p_r\!+\!1)_{2j}}{(p_r\!+\!1)_{j}}
\Pi_{s=1}^{P\!-\!K} \frac{(l_s\!+\!1)_j}{(2 l_s\!+\!1)_{2j}}
\Delta \; , 
\label{hyper}
\end{eqnarray}
\vspace{-5mm}
with
\begin{eqnarray}
\Delta  & = &   
\exp \Biggl[ \sum_{k=1}^{\infty} \frac{(-\ep)^k}{k}  
\Biggl( 
\sum_{\omega=1}^K b_\omega^k S_k(n_\omega+j-1)
\!-\! \sum_{i=1}^J a_i^k S_k(m_i \!+\! j \!-\! 1)
\nonumber \\ && \hspace{5mm}
+ \sum_{s=1}^{P-K} c_s^k \left[ S_k(2 l_s \!+\! 2j) \!-\! S_k(l_s \!+\! j) \right]
\!-\! \sum_{r=1}^{P+1-J} d_r^k \left[ S_k(2 p_r \!+\! 2j) \!-\! S_k(p_r \!+\! j) \right]
\Biggr) 
\Biggr] \; . 
\nonumber 
\end{eqnarray}
In this way, the $\ep$-expansion of the hypergeometric function 
(\ref{hyper}) is reduced to the calculation of the {\it multiple sums}
\begin{eqnarray}
&& \hspace{-7mm}
\sum_{j=1}^\infty \frac{z^j}{j!} \frac{1}{4^{j(K\!-\!J\!+\!1)}} 
\frac{ \Pi_{i=1}^J (m_i-1+j)! \Pi_{r=1}^{P+1-J} (2 p_r+2j)!}
     { \Pi_{\omega=1}^K (n_\omega-1+j)! \Pi_{s=1}^{P-K} (2 l_s+2j)!} 
\nonumber \\[-2mm] && \hspace{-5mm}
\times [S_{a_1}(m_1\!+\!j\!-\!1)]^{i_1}\ldots [S_{a_\mu}(m_\mu\!+\!j\!-\!1)]^{i_p}\; 
       [S_{b_1}(2p_r\!+\!2j)]^{j_1}\ldots [S_{b_\nu}(2p_\nu\!+\!2j)]^{j_q},
\label{multiple_sums}
\end{eqnarray}
where $m_j,n_k,l_\omega,p_r$ are positive integer numbers and $|z|<1$. 
For $z$ outside of circle of convergence it is necessary to transform 
$z$ in such a way that the original hypergeometric functions can be expressed in terms 
of other hypergeometric functions with convergent series \cite{hyper:analytical}.
However, very often it is necessary to construct the $\ep$-expansion for 
integrals of the following type \cite{expansion}
\begin{eqnarray}
\int_0^z 
\frac{dz z^\alpha}{(1-z)^\beta}
~_{2}F_1\left(\begin{array}{c|}
A, \; B \\
C 
\end{array} ~  \mu z \right) \;,
\label{two:1}
\end{eqnarray}
or 
\begin{eqnarray}
\int_0^z 
\frac{dz z^\alpha}{(1-z)^\beta}
~_{2}F_1\left(\begin{array}{c|}
A, \; B \\
C 
\end{array} ~  \mu z \right)
~_{2}F_1\left(\begin{array}{c|}
M, \; N \\
K 
\end{array} ~  \sigma z \right) \;,
\label{two:2}
\end{eqnarray}
where all parameters, in general, depend on the $\ep$.
In this case, it is more convenient to construct firstly the 
$\ep$-expansion of the hypergeometric function and then integrate it with the proper 
kernel. Moreover, by help of the representation
\begin{equation}
{}_{P+1}F_P \left( \begin{array}{c|}
a, \{A\} \\
b, \{B\}
\end{array}~ z \right)
= 
\frac{\Gamma(b)}{\Gamma(a)\Gamma(b-a)}
\int_0^1 t^{a-1} (1-t)^{b-a-1} 
{}_{P}F_{P-1} \left( \begin{array}{c|}
\{A\} \\
\{B\}
\end{array}~ t z \right) \;, 
\end{equation}
the $\ep$-expansion of any generalized hypergeometric 
function $_PF_{P-1}$ can be constructed via the $\ep$-expansion of a Gauss one. 

An algorithm for the construction of $\ep$-expansion of hypergeometric functions with integer values 
of parameters has been proposed in  \cite{nested} and has been generalized recently 
for rational values in \cite{rational}. The resulting expansion are expressible in terms 
of nested sums or multiple polylogarithms \cite{goncharov}. 

In contrast to this approach, in paper \cite{DK04} an alternative algorithm 
has been invented. It is based on construction of $\ep$-expansion of a basis of 
hypergeometric functions~\footnote{ 
The idea of a basis of hypergeometric functions is closely related to 
idea of master-integrals in high-energy perturbative calculations. Using 
algebraic relations between Feynman integrals derived by help of 
integration by parts approach \cite{rec} and/or shifting of the space-time 
dimension \cite{tarasov}, the original set of physical amplitudes can be reduced 
to a restricted set of so-called ``master-integrals''. 
For example, such reduction is the necessary step in  proving gauge invariance of 
pole masses (for recent results, see \cite{pole}). 
The algorithm of reduction is called the solution of recurrence relations.}. 
For hypergeometric functions with integer or half-integer values of parameters, 
the following basis has been analyzed:
\begin{eqnarray}
&& \hspace*{-7mm}
_{P+1}F_P\left(\begin{array}{c|}
\{ \tfrac{3}{2} +b_i\ep\}^{J}, \;
\{ 1+a_i\ep\}^K, \; \{ 2+d_i\ep\}^L  \\
\{ \tfrac{3}{2} + f_i\ep\}^{J-1}, \;
\{ 1+e_i\ep \}^R,
\{ 2+c_i\ep \}^{K+L-R}
\end{array} ~  z \right) \; , 
%%%%%%%%%%%%%%%%%%%%%%%%%%%%%%%%%%%%%%%%%%%%%%%%%%%%%%%%%%%%%%%%%%%
\\ && \hspace*{-7mm}
_{P+1}F_P\left(\begin{array}{c|}
\{ \tfrac{3}{2} +b_i\ep\}^{J}, \;
\{ 1+a_i \ep\}^K, \; \{ 2+d_i\ep\}^L  \\
\{ \tfrac{3}{2} + f_i\ep\}^{J}, \;
\{ 1+e_i\ep \}^R,
\{ 2+c_i\ep \}^{K+L-R-1}
\end{array} ~  z \right) \; , 
%%%%%%%%%%%%%%%%%%%%%%%%%%%%%%%%%%%%%%%%%%%%%%%%%%%%%%%%%%%%%%%%%%%
\\ && \hspace*{-7mm}
_{P+1}F_P\left(\begin{array}{c|}
\{ \tfrac{3}{2} +b_i\ep\}^{J-1}, \; 
\{ 1+a_i\ep\}^K, \; \{ 2+d_i\ep\}^L  \\
\{ \tfrac{3}{2} + f_i\ep\}^J, \; 
\{ 1+e_i\ep \}^R,
\{ 2+c_i\ep \}^{K+L-R-2} 
\end{array} ~z \right) \;.
\label{basis}
\end{eqnarray}
In this case, the sums (\ref{multiple_sums})
are reduced to {\it multiple sums} of the following form, 
\begin{eqnarray}
&& 
\Sigma^{(k)}_{a_1,\ldots,a_p; \; b_1,\ldots,b_q;c}(u)
\equiv
\sum_{j=1}^\infty \tfrac{1}{\left( 2j \atop j\right)^k }\frac{u^j}{j^c}
S_{a_1} \ldots S_{a_p} \bar{S}_{b_1} \ldots \bar{S}_{b_q} \; , 
\label{binsum}
\end{eqnarray}
where $u$ is, in general, an arbitrary argument (in this particular case it is equal to $4^k z$)
and we accept that the notations $S_a$ and $\bar{S}_b$
will always mean $S_a(j-1)$ and $S_b(2j-1)$, respectively.
For particular values of $k$, the sums (\ref{binsum}) are called 
\begin{eqnarray}
k = 
\left\{ 
\begin{array}{cl}
 0  & \mbox{ {\it harmonic} } \\
 1  & \mbox{ {\it inverse binomial} } \\
-1  & \mbox{ {\it binomial} } 
\end{array} \right\} \mbox{ sums }
\nonumber 
\end{eqnarray}
The analytical results for harmonic, binomial and inverse binomial sums 
of different weights and depths have been presented in \cite{harmonic,FKV,DK04},
\cite{binomial,DK04}, \cite{ibs,DK04}, respectively.
The results of the $\ep$-expansion are expressed in terms of 
polylogarithms \cite{Lewin}, Nielsen polylogarithms \cite{Nielsen} or 
harmonic polylogarithms \cite{RV00}.
The missing part of the approach described in \cite{DK04} is an algorithm for the reduction 
of original functions to our basis (\ref{basis}).
However, for all physically important cases \cite{physical_example,particular}, 
the solutions have been presented. They were derived as the solution of 
recurrence relations for the proper Feynman 
diagrams \cite{programs:feynman,programs:propagator}.

In this paper we construct an algorithm for the reduction of 
a Gauss hypergeometric function with arbitrary parameters to 
two Gauss hypergeometric functions with defined parameters
(reduction to the master-integrals). For integer and half-integer values 
of parameters~\footnote{Unfortunately, in physical applications other
values of parameters also exist \cite{quarter}.}, 
the $\ep$-expansion is constructed up to functions of {\it weight} {\bf 4}. 
As an illustration of the elaborated algorithm, some multiloop scalar 
integrals are calculated. We like to note that 
for some physically important cases, the proper $\ep$-expansions 
of Gauss hypergeometric functions have been presented in 
\cite{BD,FJTV,DK01,one-loop:1,one-loop:2}.
The all-order $\ep$-expansion of basis Gauss hypergeometric functions 
with integer values of parameters is constructed in \cite{hyper:expansion}.
The results are expressible in terms of {\it multiple polylogarithms}.
The algebraic Gauss hypergeometric functions have been studied in 
\cite{hyper:algebraic}.
The numerical approaches are discussed in \cite{hyper:numerical}.

%%%%%%%%%%%%%%%%%%%%%%%%%%%%%%%%%%%%%%%%%%%%%%%%%%%%%%%%%%%%%%%%%%%%%%%%%
%%%%%%%%%%%%%%%%%%%%%%%%%%%%%%%%%%%%%%%%%%%%%%%%%%%%%%%%%%%%%%%%%%%%%%%%
\section{Reduction of Gauss hypergeometric functions}
\setcounter{equation}{0}
As is known, for any three contiguous Gauss hypergeometric functions 
there is a contiguous relation, which is a linear relation
with coefficients being rational functions in the parameters
$A,B,C$ and argument $z$. Using the well know relations \cite{bateman}

\begin{eqnarray}
&& 
\hspace{-10mm}
A (1-z)
~_{2}F_1\left(\begin{array}{c|}
A+1, B \\
C \end{array} ~z \right) 
- 
(C-A)
~_{2}F_1\left(\begin{array}{c|}
A-1, B \\ 
C \end{array} ~z \right) 
\nonumber \\ && \hspace{10mm}
= 
\left [ 2A\!-\!C\!-\!(A\!-\!B)z \right]
~_{2}F_1\left(\begin{array}{c|}
A, B \\
C \end{array} ~z \right) \; ,
\label{recI:1}
\\ && \hspace{-10mm}
%%%%%%%%%%%%%%%%%%%%%%%%%%%%%%%%%%
(C-A) (C-B) z
~_{2}F_1\left(\begin{array}{c|}
A, B \\
C+1 \end{array} ~z \right)
- 
C(C\!-\!1)(1\!-\!z)
~_{2}F_1\left(\begin{array}{c|}
A, B \\ 
C-1 \end{array} ~z \right) 
\nonumber \\ && \hspace{10mm}
= 
C \left[1 \!-\! C\!-\!(1\!+\!A\!+\!B\!-\!2C)z \right] 
~_{2}F_1\left(\begin{array}{c|}
A, B \\
C \end{array} ~z \right) \; , 
\label{recI:2}
\end{eqnarray}
any Gauss hypergeometric function with arbitrary parameters is reduced to the combination of 
eight ones
$$
~_{2}F_1\left(\begin{array}{c|}
\{a,a+1\}, \{b,b+1\} \\
\{c,c+1\} \end{array} ~z \right)\; , 
$$
where $a,b,c$ are some fixed values of parameters.
Applying the relations 
%%%%%%%%%%%%%%%%%%%%%%%%%%%%%%%%%%
\begin{eqnarray}
&&  \hspace{-10mm}
(c-a) (c-b)
~_{2}F_1\left(\begin{array}{c|}
a, b \\
c+1 \end{array} ~z \right)
= 
\nonumber \\ && \hspace{10mm}
ab(1\!-\!z)
~_{2}F_1\left(\begin{array}{c|}
a+1, b+1 \\ 
c+1 \end{array} ~z \right) 
- 
c(a \!+\! b\!-\!c) 
~_{2}F_1\left(\begin{array}{c|}
a, b \\
c \end{array} ~z \right) \; , 
\label{recII:1}
\\ && \hspace{-10mm}
%%%%%%%%%%%%%%%%%%%%%%%%%%%%%%%%%%
c
~_{2}F_1\left(\begin{array}{c|}
a+1, b \\
c \end{array} ~z \right)
= 
bz
~_{2}F_1\left(\begin{array}{c|}
a+1, b+1 \\ 
c+1 \end{array} ~z \right) 
+ 
c
~_{2}F_1\left(\begin{array}{c|}
a, b \\
c \end{array} ~z \right) \; , 
\label{recII:2}
\\ && \hspace{-10mm}
%%%%%%%%%%%%%%%%%%%%%%%%%%%%%%%%%%
c(1-z)
~_{2}F_1\left(\begin{array}{c|}
a+1, b+1 \\
c \end{array} ~z \right)
= 
\nonumber \\ && \hspace{10mm}
(1\!+\!a\!+\!b\!-\!c)z
~_{2}F_1\left(\begin{array}{c|}
a+1, b+1 \\ 
c+1 \end{array} ~z \right) 
+ 
c
~_{2}F_1\left(\begin{array}{c|}
a, b \\
c \end{array} ~z \right) \; , 
\label{recII:3}
\\ && \hspace{-10mm}
%%%%%%%%%%%%%%%%%%%%%%%%%%%%%%%%%%
(b-c)
~_{2}F_1\left(\begin{array}{c|}
a+1, b \\
c+1 \end{array} ~z \right)
= 
b (1-z)
~_{2}F_1\left(\begin{array}{c|}
a+1, b+1 \\ 
c+1 \end{array} ~z \right) 
- 
c
~_{2}F_1\left(\begin{array}{c|}
a, b \\
c \end{array} ~z \right) \; , 
%%%%%%%%%%%%%%%%%%%%%%%%%%%%%%%%%%
\label{recII:4}
\end{eqnarray}
we are able to reduce an original Gauss hypergeometric function 
to the linear combination of two (our basis)
\begin{eqnarray}
_{2}F_1\left(\begin{array}{c|}
a, b\\
c  \end{array} ~z \right) \; , 
\quad 
_{2}F_1\left(\begin{array}{c|}
a\!+\!1, b\!+\!1 \\
c\!+\!1 \end{array} ~z \right) \; .
\label{basis-function}
\end{eqnarray}
These basis functions (\ref{basis-function}) are related by a differential identity:
\begin{eqnarray}
\frac{d}{d z}
~_{2}F_1\left(\begin{array}{c|}
a, b\\
c \end{array} ~z \right)
= \frac{ab}{c} 
~_{2}F_1\left(\begin{array}{c|}
a+1, b+1\\
c+1 \end{array} ~z \right) \; . 
\label{differential}
\end{eqnarray}
In the case when some of the parameters are positive integers (let us put $B=m$),
after applying the relation (\ref{recI:1}) we get 
one function with the value of one of the parameter equal to unity and some polynomial 
with respect to $z$ (parameter $B=0$).
In this case, instead of relations Eqs.~(\ref{recII:1})-(\ref{recII:4}) 
the following two relations (see \cite{bateman}) should be used for further reduction:
\begin{eqnarray}
&& 
\hspace{-10mm}
a (1-z)
~_{2}F_1\left(\begin{array}{c|}
1,a+1 \\
c \end{array} ~z \right) 
= 
(c-1)
+ 
(1\!+\!a\!-\!c) 
~_{2}F_1\left(\begin{array}{c|}
1, a \\
c \end{array} ~z \right) \; ,
\label{recIII:1}
\\ && \hspace{-10mm}
%%%%%%%%%%%%%%%%%%%%%%%%%%%%%%%%%%
(a-c) z
~_{2}F_1\left(\begin{array}{c|}
1, a \\
c+1 \end{array} ~z \right)
= 
- c \Biggl[
1 - (1-z)
~_{2}F_1\left(\begin{array}{c|}
1, a \\ 
c \end{array} ~z \right) 
\Biggr ] \; .
\label{recIII:2}
\end{eqnarray}
In this way, if one of the upper parameters is an integer, then the result
of reduction is expressible in terms of one Gauss hypergeometric function 
and a polynomials (the function ${}_1F_0$).
For case $c=b$, the relations (\ref{recII:1}) and (\ref{recII:4}) are useless (0=0). 
In this case, we should apply the Kummer relation (\ref{Kummer:1}) or (\ref{Kummer:2}) 
and reduce this case to the previous one (\ref{recIII:1}) and (\ref{recIII:2}):

\begin{eqnarray}
~_{2}F_1\left(\begin{array}{c|}
A,b \\
1+b \end{array} ~z \right) 
= 
\frac{1}{(1\!-\!z)^A}
~_{2}F_1\left(\begin{array}{c|}
1, A \\
1\!+\!b \end{array} ~ - \frac{z}{1\!-\!z} \right) 
= 
(1\!-\!z)^{1\!-\!A}
~_{2}F_1\left(\begin{array}{c|}
1, 1\!+\!b\!-\!A \\
1+b \end{array} ~ z \right) \;.
\end{eqnarray}
%%%%%%%%%%%%%%%%%%%%%%%%%%%%%%%%%%%%%%%%%%%%%%%%%%%%%%%%
Another algorithm of reduction is described in \cite{hyper:contiguous}.

%%%%%%%%%%%%%%%%%%%%%%%%%%%%%%%%%%%%%%%%%%%%%%%%%%%%%%%%%%%%%%%%%%%%%%%%%%%%%%%%%%%%%%%%%
\section{Relations between basis hypergeometric functions with integer or half-integer 
          values of parameters}
\label{set}
Let us consider a Gauss hypergeometric functions with 
integer or half-integer values of $\ep$-independent parameters. 
In this case, the set of basis functions consist of the 12 (sixth time two) functions.
We will call these basis functions as functions of type {\bf A, B, C, D, E, F}.
For each type the values of $a,b,c$, parameters of our basis (\ref{basis-function}), are presented in Table I:
%%%%%%%%%%%%%%%%%%%%%%%%%%%%%%%%%%%%%%%%%%%%%%%%%%%%%%%%
$$
\begin{array}{|c|c|c|c|c|c|c|} \hline
\multicolumn{7}{|c|}{Table ~~~I }    \\   \hline
      & {\bf A}             & {\bf B}             & {\bf C}             & {\bf D}              & {\bf E} & 
{\bf F} \\[0.3cm] \hline
a     & a_1 \ep             & a \ep               & a \ep           & \frac{1}{2} + b_1 \ep & a_1 \ep   & 
\frac{1}{2} + b_1 \ep  \\[0.3cm] \hline
b     & a_2 \ep             & \frac{1}{2} + b \ep & \frac{1}{2} + b \ep & \frac{1}{2} + b_2 \ep & a_2 \ep   & 
\frac{1}{2} + b_2 \ep  \\[0.3cm] \hline
c     & \frac{1}{2} + f \ep & 1 + c \ep           & \frac{1}{2} + f \ep & \frac{1}{2} + f \ep   & 1 + c \ep & 
1 + c \ep 
\\[0.3cm] \hline
\end{array}
$$
%%%%%%%%%%%%%%%%%%%%%%%%%%%%%%%%%%%%%%%%%%%%%%%%%%%%%%%%
The number of independent basis hypergeometric functions, enumerated 
in Table I, can be reduced  by help of the Kummer transformations \cite{hyper:kummer} of variable $z$. 
With respect to this transformations the functions of type {\bf A, B, C, D} are 
transformed into each other. 
This allows us to reduce the number of independent hypergeometric functions.
The functions of type {\bf E, F} transform into functions of the same type. 
%%%%%%%%%%%%%%%%%%%%%%%%%%%%%
Let us illustrate how  functions of type {\bf B, C, D} can be expressed in 
terms of functions of type {\bf A}.
Starting from the relation 
\begin{eqnarray}
\left. _2F_1\left( \begin{array}{c} 
a,b\\
c \end{array} \right| z  \right) 
& =  & (1-z)^{c-a-b}
\left. _2F_1\left( \begin{array}{c} 
c-a, c-b \\
c \end{array} \right| z \right) \; , 
\label{Kummer:1}
\end{eqnarray}
we express the functions of ${\bf D}$-type in terms of functions of {\bf A}-type
%%%%%%%%%%%%%%%%%%%%%%%%%%%%%
%%%%%%%%%%%%%%%%%%%%%%%%%%%%%
\begin{eqnarray}
\left. _2F_1\left( \begin{array}{c} 
\tfrac{1}{2} + b_1 \ep, \tfrac{1}{2} + b_2 \ep\\
\tfrac{1}{2} + f \ep \end{array} \right| z  \right) 
& = &  \frac{(1-z)^{(f-b_1-b_2)\ep}}{(1-z)^{1/2}}
\left. _2F_1\left( \begin{array}{c} 
(f \!-\! b_1)\ep, (f \!-\! b_2) \ep \\
\tfrac{1}{2} \!+\! f \ep 
\end{array} \right| z \right) \; , 
%%%%%%%%%%%%%%%%%%%%%%%%%%%%%%%%%%%%%%%%%%%%%
\label{D->A:1}
\\ 
\left. _2F_1\left( \begin{array}{c} 
\tfrac{3}{2} \!+\! b_1 \ep, \tfrac{3}{2} \!+\! b_2 \ep\\
\tfrac{3}{2} \!+\! f \ep \end{array} \right| z  \right) 
& = &  
\frac{(1\!-\!z)^{(f\!-\!b_1\!-\!b_2)\ep}}{(1\!-\!z)^{3/2} (1\!+\!2b_1\ep)(1\!+\!2b_2\ep)}
\times
\nonumber \\ && \hspace{-50mm}
\Biggl\{ 
4 (1 \!-\! z) (f\!-\!b_1)(f\!-\!b_2) \ep^2 
\left. _2F_1\left( \begin{array}{c} 
1 \!+\! (f \!-\! b_1)\ep, 1 \!+\! (f \!-\! b_2) \ep \\
\tfrac{3}{2} \!+\! f \ep 
\end{array} \right| z \right) 
 \nonumber \\ && \hspace{-50mm}
+ 
(1\!+\!2f\ep) [1\!-\!2(f\!-\!b_1\!-\!b_2)\ep]
\left. _2F_1\left( \begin{array}{c} 
(f \!-\! b_1)\ep, (f \!-\! b_2) \ep \\
\tfrac{1}{2} \!+\! f \ep 
\end{array} \right| z \right) 
\Biggr\} \; , 
\label{D->A:2}
\end{eqnarray}
%%%%%%%%%%%%%%%%%%%%%%%%%%%%%
Functions of {\bf C}-type can be written as a linear combination of functions of {\bf A}-type.
Using relation
%%%%%%%%%%%%%%%%%%%%%%%%%%%%%
\begin{eqnarray}
\left. _2F_1\left( \begin{array}{c} 
a,b\\
c \end{array} \right| z  \right) 
& = & \frac{1}{(1-z)^a}
\left. _2F_1\left( \begin{array}{c} 
a, c-b \\
c \end{array} \right| - \frac{z}{1-z} \right) \; , 
\label{Kummer:2}
\end{eqnarray}
we get 
\begin{eqnarray}
&& 
\left. _2F_1\left( \begin{array}{c} 
\tfrac{1}{2} \!+\! b, a \ep \\
\tfrac{1}{2} \!+\! f \ep 
\end{array} \right| z \right) 
=   
\frac{1}{(1-z)^{a \ep}}
\left. _2F_1\left( \begin{array}{c} 
a \ep, (f\!-\!b) \ep\\
\tfrac{1}{2} \!+\! f \ep \end{array} \right| - \frac{z}{1-z}   \right) 
\; , 
%%%%%%%%%%%%%%%%%%%%%%%%%%%%%%%%%%%%%%%%%%%%%
\label{C->A:1}
\\ && 
(1 \!+\! 2b\ep) (1-z)^{1+a\ep}
\left. _2F_1\left( \begin{array}{c} 
1 \!+\! a \ep, \tfrac{3}{2}\!+\! b \ep\\
\tfrac{3}{2} \!+\! f \ep \end{array} \right| z \right) 
=  
(1\!+\! 2 f \ep)
\left. _2F_1\left( \begin{array}{c} 
(f\!-\!b) \ep, a \ep \\
\tfrac{1}{2} \!+\! f \ep 
\end{array} \right| - \frac{z}{1-z}  \right) 
\nonumber \\ && \hspace{10mm}
- \frac{2(f\!-\!b)\ep }{(1\!-\!z)}
\left. _2F_1\left( \begin{array}{c} 
1\!+\! (f \!-\! b)\ep, 1 \!+\! a \ep \\
\tfrac{3}{2} \!+\! f \ep 
\end{array} \right| - \frac{z}{1-z}  \right) 
\Biggr\} \; .
\label{C->A:2}
\end{eqnarray}
%
%
%%%%%%%%%%%%%%%%%%%%%%%%%%%%%
Using the transformation $z \to 1 - \frac{1}{z}$, 
\begin{eqnarray}
\left. _2F_1\left( \begin{array}{c} 
a,b\\
c \end{array} \right| z  \right) 
& =  & 
\frac{1}{z^a} \frac{\Gamma(c) \Gamma(c-a-b)}{\Gamma(c-a)\Gamma(c-b)}
\left. _2F_1\left( \begin{array}{c} 
a, 1+a-c \\
1+a+b-c \end{array} \right| 1 - \frac{1}{z} \right) 
\nonumber \\ 
& + & 
z^{a-c} (1\!-\!z)^{c-a-b}
\frac{\Gamma(c) \Gamma(a+b-c)}{\Gamma(a)\Gamma(b)}
\left. _2F_1\left( \begin{array}{c} 
c\!-\!a, 1\!-\!a \\
1\!+\!c\!-\!a\!-\!b \end{array} \right| 1 \!-\! \frac{1}{z} \right) \;,
\label{1-1/z}
\end{eqnarray}
all functions of {\bf B}-type can be presented as a linear combination of functions of {\bf A}-type
\begin{eqnarray}
&& \hspace{-5mm}
\left. _2F_1\left( \begin{array}{c} 
\tfrac{3}{2} \!+\! b\ep, 1\!+\!a \ep \\
2 \!+\! c \ep 
\end{array} \right| z \right) 
=   
\frac{1}{z^{1+a\ep}}
\frac{\Gamma(2\!+\!c\ep) \Gamma\left(-\frac{1}{2} \!+\! (c\!-\!a\!-\!b)\ep \right)}
     {\Gamma(1\!+\!(c\!-\!a)\ep) \Gamma\left(\frac{1}{2} \!+\! (c\!-\!b)\ep \right) (1\!+\!2b\ep)}
\nonumber \\ && 
\times
\Biggl\{
\left[ 
1 + 2 (a+b-c)\ep 
\right]
\left. _2F_1\left( \begin{array}{c} 
a \ep, (a-c) \ep\\
\tfrac{1}{2} \!+\! (a+b-c)\ep \end{array} \right| 1 \!-\! \frac{1}{z} \right) 
\nonumber \\ && \hspace{30mm}
-
2 \frac{(a-c)\ep}{z}
\left. _2F_1\left( \begin{array}{c} 
1 \!+\! (a\!-\!c) \ep, 1 \!+\! a \ep \\
\tfrac{3}{2} \!+\! (a\!+\!b\!-\!c) \ep 
\end{array} \right| 1 \!-\! \frac{1}{z}  \right) 
\Biggr\} 
\nonumber \\ && 
- 
\frac{\Gamma(2\!+\!c\ep) \Gamma\left(-\frac{1}{2} \!-\! (c\!-\!a\!-\!b)\ep \right)}
     {\Gamma(1 \!+\! a\ep) \Gamma\left(\frac{3}{2} \!+\! b\ep \right)}
\frac{z^{-1+(a-c)\ep}}{(1-z)^{1/2-(c-a-b)\ep}}
\nonumber \\ && 
\times
\Biggl\{
\frac{a \ep (1-z)}{z}
\left. _2F_1\left( \begin{array}{c} 
1\!+\! (c \!-\! a)\ep, 1 \!-\! a \ep \\
\tfrac{3}{2} \!+\! (c\!-\!a\!-\!b) \ep 
\end{array} \right| 1 - \frac{1}{z}  \right) 
\nonumber \\ && \hspace{30mm}
+ 
\left[ 
\frac{1}{2} \!+\! (c\!-\!a\!-\!b)\ep
\right]
\left. _2F_1\left( \begin{array}{c} 
(c \!-\! a)\ep, - a \ep \\
\tfrac{1}{2} \!+\! (c\!-\!a\!-\!b) \ep 
\end{array} \right| 1 - \frac{1}{z}  \right) 
\Biggr\} \; ,
\label{B->A:1}
%%%%%%%%%%%%%%%%%%%%%%%%%%%%%%%%%%%%%%%%%%%%%%%%%%%%
\\ && \hspace{-5mm}
\left. _2F_1\left( \begin{array}{c} 
\tfrac{1}{2} \!+\! b\ep, a \ep \\
1 \!+\! c \ep 
\end{array} \right| z \right) 
\nonumber \\ && 
=   
\frac{\Gamma(1\!+\!c\ep) \Gamma\left(-\frac{1}{2} \!-\! (c-a-b)\ep \right)}
     {\Gamma(a\ep) \Gamma\left(\frac{1}{2} + b\ep \right)}
\frac{(1-z)^{1/2+(c-a-b)\ep}}{z^{1-(a-c)\ep}}
\left. _2F_1\left( \begin{array}{c} 
1+(c \!-\! a)\ep, 1\!-\! a \ep \\
\tfrac{3}{2} \!+\! (c\!-\!a\!-\!b) \ep 
\end{array} \right| 1 - \frac{1}{z}  \right) 
\nonumber \\ && 
+ 
\frac{1}{z^{a\ep}}
\frac{\Gamma(1\!+\!c\ep) \Gamma\left(\frac{1}{2} \!+\! (c\!-\!a\!-\!b)\ep \right)}
     {\Gamma(1+(c-a)\ep) \Gamma\left(\frac{1}{2} \!+\! (c\!-\!b)\ep \right)}
\left. _2F_1\left( \begin{array}{c} 
a \ep, (a-c) \ep\\
\tfrac{1}{2} \!+\! (a+b-c)\ep \end{array} \right| 1 - \frac{1}{z} \right) 
\; .
\label{B->A:2}
\end{eqnarray}
As a result, we get the following statement:

%%%%%%%%%%%%%%%%%%%%%%%%%%%%%%%%%%%%%%%%%%%%%%%%%%%%%%%%
\noindent
{\it 
Any functions of type {\bf A, B, C, D} can be expressed in an algebraic way 
in terms of just one of these types.
}
%%%%%%%%%%%%%%%%%%%%%%%%%%%%%%%%%%%%%%%%%%%%%%%%%%%%%%%

\noindent
By help of the representation \cite{bateman},
\begin{eqnarray}
\left. _2F_1\left( \begin{array}{c} 
a, b \\
c \end{array} \right| z\right) 
= \frac{\Gamma(c)}{\Gamma(m) \Gamma(c-m)}
\int_0^1 
dx x^{m-1} (1-x)^{c-m-1}
\left. _2F_1\left( \begin{array}{c} 
a, b \\
m \end{array} \right| xz \right) \; , 
\label{integral_representation}
\end{eqnarray}
it is possible to find the integral relations between the coefficients of 
$\ep$-expansion of basis functions. In our case (integer and half-integer values of parameters) 
there are integral relations between functions of type 
${\bf A}$ and ${\bf E}$; 
${\bf B}$ and ${\bf C}$; 
${\bf D}$ and ${\bf F}$. 

Putting $m=1/2$ in the r.h.s. of Eq.(\ref{integral_representation}) and 
using the quadratic transformation \cite{bateman}
\begin{eqnarray}
&& 
\frac{2 \Gamma\left(\frac{1}{2}\right) \Gamma\left(a\!+\!b\!+\!\frac{1}{2}\right)}
     {\Gamma\left(a\!+\!\frac{1}{2}\right) \Gamma\left(b\!+\!\frac{1}{2}\right)}
\left. _2F_1\left( \begin{array}{c} 
a, b \\
\frac{1}{2} \end{array} \right| z \right) 
\nonumber \\ && \hspace{20mm}
= 
\left. _2F_1\left( \begin{array}{c} 
2a, 2b \\
a\!+\!b\!+\!\frac{1}{2} \end{array} \right| \frac{1\!+\!\sqrt{z}}{2} \right) 
+
\left. _2F_1\left( \begin{array}{c} 
2a, 2b \\
a\!+\!b\!+\!\frac{1}{2} \end{array} \right| \frac{1\!-\!\sqrt{z}}{2} \right) 
\; , 
\end{eqnarray}
we get the representation
\begin{eqnarray}
&& \hspace{-5mm}
\left. _2F_1\left( \begin{array}{c} 
a, b \\
c \end{array} \right| z\right) 
= \frac{\Gamma(c) \Gamma\left(a\!+\!\tfrac{1}{2}\right) \Gamma\left(b\!+\!\tfrac{1}{2}\right)}
       {2 \pi \Gamma\left(c-\tfrac{1}{2}\right) \Gamma\left(a\!+\!b\!+\!\tfrac{1}{2}\right)}
\nonumber \\ && 
\times
\int_0^1 
\frac{dx}{\sqrt{x}} (1-x)^{c-3/2}
\Biggl[
\left. _2F_1\left( \begin{array}{c} 
2a, 2b \\
a\!+\!b\!+\!\frac{1}{2} \end{array} \right| \frac{1\!+\!\sqrt{xz}}{2} \right) 
\!+\!
\left. _2F_1\left( \begin{array}{c} 
2a, 2b \\
a\!+\!b\!+\!\frac{1}{2} \end{array} \right| \frac{1\!-\!\sqrt{xz}}{2} \right) 
\Biggr]
\; .
\label{integral_representationII}
\end{eqnarray}
This representation allows us to find integral relation between coefficients of 
the $\ep$-expansion of functions of ${\bf B}$ and ${\bf E}$ types; 
${\bf C}$ and ${\bf E}$ types; 
and ${\bf F}$ and ${\bf A}$ types.
%%%%%%%%%%%%%%%%%%%%%%%%%%%%%%%%%%%%%%%%%%%%%%%%%%%%%%%%
%%%%%%%%%%%%%%%%%%%%%%%%%%%%%%%%%%%%%%%%%%%%%%%%%%%%%%%%%%%%%%%%%%%%%%%%%%%%%%%%%%%%%%%%%
\section{Laurent expansion of basis functions with integer or half-integer 
            values of parameters}

In accordance with the algorithm described in \cite{DK04}, the 
Laurent expansion of our basis functions (see Table~I in section \ref{set}) with respect to parameter $\ep$
is reduced to the study of multiple series of type (\ref{multiple_sums}).
Using the general expressions given in \cite{DK04} (see Eqs.(2.30), (2.31), (D.1), (D.2) in \cite{DK04})
we derive that hypergeometric functions of type ${\bf B}$
and ${\bf D}$ are expressible in terms of {\it multiple binomial sums} and 
hypergeometric functions of type ${\bf C}$
and ${\bf E}$ are expressible in terms of {\it multiple harmonic sums}.
The hypergeometric functions of type ${\bf A}$
and ${\bf E}$ are expressible in terms of {\it multiple inverse binomial sums}.
The functions of type {\bf F} include {\it multiple double binomial sums} (see Eq.~(\ref{double})).

Here we present the $\ep$-expansion of our basis functions, enumerated in Table~I.
We restrict ourself by constructing the $\ep$-expansion up to functions of {\it weight} {\bf 4}. 
We like to note that our expansion is organised in such manner
that a term $O(\ep^k)$ in brackets means functions of {\it weight} {\bf 5}. 
At this order the Nielsen polylogarithms are not enough for constructing 
the $\ep$-expansion and new function (one at least) should be introduced. 

\subsection{type {\bf A}}
The $\ep$-expansion of functions of type {\bf A} have been studied in our previous paper \cite{DK04}.
Here, we collect the proper results (see Eqs.~(2.3), (2.30) and Table I in Appendix C):

\begin{eqnarray}
&& \hspace{-5mm}
\left. _2F_1\left( \begin{array}{c} 1+a_1 \ep, 1 + a_2 \ep \\
                 \frac{3}{2} + f \ep \end{array} \right| z\right) 
= 
\frac{(1+2f\ep)}{2z} \frac{1-y}{1+y}
\Biggl(
\ln y 
\nonumber \\ && \hspace{-3mm}
+ \ep \Biggl\{
2 (f\!-\!a_1\!-\!a_2) \left[ \Li{2}{-y} \!+\! \ln y \ln (1+y) \right]
- 2 f \left[ \Li{2}{y} \!+\! \ln y \ln (1-y) \right]
\nonumber \\ && \hspace{7mm}
+ \frac{1}{2} (a_1+a_2) \ln^2 y 
+ \zeta_2 (3f-a_1-a_2)
\Biggr\}
\nonumber \\ && \hspace{-3mm}
+ \ep^2 \Biggl\{
 4 (a_1\!+\!a_2\!-\!f) (a_1\!+\!a_2\!-\!2f) \Snp{1,2}{-y}
\nonumber \\ && \hspace{7mm}
\!-\! 4 f (a_1\!+\!a_2\!-\!2f) \Snp{1,2}{y}
\!+\! 2 f (a_1\!+\!a_2\!-\!f) \Snp{1,2}{y^2}
\nonumber \\ && \hspace{7mm}
- 2 (a_1\!+\!a_2) (a_1\!+\!a_2\!-\!f) \Li{3}{-y}
- 2 f (a_1\!+\!a_2) \Li{3}{y}
+ 4 f^2 \ln(1-y) \Li{2}{y}
\nonumber \\ && \hspace{7mm}
+ 4 f (a_1\!+\!a_2\!-\!f) \left[ \ln (1\!+\!y )\Li{2}{y}
\!+\! \ln (1\!-\!y) \Li{2}{-y} \right]
\nonumber \\ && \hspace{7mm}
+ 2 (a_1\!+\!a_2\!-\!f)^2 \ln (1+y) \left[ 2\Li{2}{-y} \!+\! \ln y \ln (1\!+\!y) \right]
+ 2 f^2 \ln y \ln^2 (1\!-\!y)
\nonumber \\ && \hspace{7mm}
- 2 (3f\!-\!a_1\!-\!a_2) \zeta_2 
% it was typo here
\left[ f \ln (1 \!-\! y) + (a_1\!+\!a_2\!-\!f) \ln (1 \!+\! y) \right]
\nonumber \\ && \hspace{7mm}
+ 4 f (a_1\!+\!a_2\!-\!f) \ln y \ln (1\!+\!y) \ln (1\!-\!y)
- f (a_1+a_2) \ln^2 y \ln (1-y)
\nonumber \\ && \hspace{7mm}
- (a_1+a_2) (a_1\!+\!a_2\!-\!f) \ln^2 y \ln (1+y)
+ (a_1+a_2) (3f\!-\!a_1\!-\!a_2) \zeta_2 \ln y 
\nonumber \\ && \hspace{7mm}
+ \frac{1}{6} (a_1^2+a_2^2+a_1a_2) \ln^3 y 
+ \zeta_3 \left[7f (a_1+a_2\!-\!f) - 2 (a_1+a_2)^2 \right]
\Biggr\}
\nonumber \\ && \hspace{-3mm}
+ \ep^3 \Biggl\{
 4 f (a_1\!+\!a_2\!-\!f) (a_1\!+\!a_2\!-\!2f) 
\left[ \Li{4}{\frac{1-y}{1+y}} \!-\! \Li{4}{-\frac{1-y}{1+y}} \right]
\nonumber \\ && \hspace{7mm}
+ 4 (a_1\!+\!a_2) (a_1\!+\!a_2\!-\!f) (a_1\!+\!a_2\!-\!2f) 
\left[ \Snp{2,2}{-y} \!-\! 2 \Snp{1,3}{-y} \!-\! 2\Snp{1,2}{-y} \ln (1\!+\!y) \right]
\nonumber \\ && \hspace{7mm}
- 4 f (a_1\!+\!a_2) (a_1\!+\!a_2\!-\!2f) 
\left[ \Snp{2,2}{y} \!-\! 2 \Snp{1,3}{y} \!-\! 2 \Snp{1,2}{y} \ln (1\!-\!y) \right]
\nonumber \\ && \hspace{7mm}
+ f (a_1\!+\!a_2) (a_1\!+\!a_2\!-\!f) 
\left[ 
\Snp{2,2}{y^2} \!-\! 2 \Snp{1,3}{y^2} \!-\! \Snp{1,2}{y^2} \ln (1\!-\!y^2) 
\right]
\nonumber \\ && \hspace{7mm}
- 2 (a_1\!-\!a_2)^2 \left[ f \Li{4}{y} \!+\! (a_1\!+\!a_2\!-\!f) \Li{4}{-y} \right]
- 4 a_1 a_2 f \ln y \Li{3}{y}
\nonumber \\ && \hspace{7mm}
+ 2 f (a_1\!+\!a_2) (a_1\!+\!a_2\!-\!f) 
\ln(1\!-\!y) \left[ 2 \Li{3}{-y} - \ln(1+y) \Li{2}{y^2} \right]
\nonumber \\ && \hspace{7mm}
+ 4 (a_1\!+\!a_2) 
\left[ 
(a_1\!+\!a_2\!-\!f)^2 \ln (1+y) \Li{3}{-y} + f^2  \ln (1-y) \Li{3}{y} \right]
\nonumber \\ && \hspace{7mm}
+ 4 (a_1+a_2-f) 
\left[
f (a_1 \!+\! a_2) \ln (1\!+\!y) \Li{3}{y} \!-\! a_1 a_2 \ln y \Li{3}{-y}
\right]
\nonumber \\ && \hspace{7mm}
- 2 (a_1\!+\!a_2) (a_1\!+\!a_2\!-\!f) (2a_1\!+\!2a_2\!-\!3f)  \ln^2 (1+y) \Li{2}{-y}
+ a_1 a_2 f \ln^2 y \Li{2}{y}
\nonumber \\ && \hspace{7mm}
- 2 f (a_1\!+\!a_2) (a_1\!+\!a_2\!-\!f) 
\left[ 
\ln^2(1\!-\!y) \Li{2}{-y} + \ln^2(1\!+\!y) \Li{2}{y}
\right]
\nonumber \\ && \hspace{7mm}
+ a_1 a_2 (a_1\!+\!a_2\!-\!f) \ln^2 y \Li{2}{-y}
+ 2 f (a_1\!+\!a_2) (a_1\!+\!a_2\!-\!3f) \ln^2 (1-y) \Li{2}{y}
\nonumber \\ && \hspace{7mm}
+ f^2 (a_1+a_2) \ln^2 y \ln^2 (1\!-\!y)
- \frac{1}{3} f (a_1^2 \!+\! a_2^2 \!+\! a_1 a_2) \ln^3 y \ln (1-y)
\nonumber \\ && \hspace{7mm}
+ (a_1 \!+\! a_2) (a_1\!+\!a_2\!-\!f) \ln^2 y \ln (1+y) 
\left[ 
(a_1\!+\!a_2\!-\!f) \ln (1+y) + 2 f  \ln (1\!-\!y) 
\right]
\nonumber \\ && \hspace{7mm}
- (a_1\!+\!a_2) (a_1\!+\!a_2\!-\!f) ( 2a_1 \!+\! 2a_2 \!-\! 5 f) \zeta_2 \ln^2 (1\!+\!y)
\nonumber \\ && \hspace{7mm}
- \frac{2}{3} (a_1\!+\!a_2) (a_1\!+\!a_2\!-\!f) ( 2a_1 \!+\! 2a_2 \!-\! 3 f) \ln y \ln^3 (1\!+\!y) 
\nonumber \\ && \hspace{7mm}
+  \frac{2}{3} f (a_1 \!+\! a_2) (a_1 \!+\! a_2 \!-\! 3 f) \ln y \ln^3 (1\!-\!y)
\nonumber \\ && \hspace{7mm}
- 2 f (a_1\!+\!a_2) ( a_1 \!+\! a_2 \!-\! f)\ln y \ln (1\!-\!y) \ln (1\!+\!y) 
\left[\ln(1\!-\!y) \!+\! \ln(1\!+\!y) \right]
\nonumber \\ && \hspace{7mm}
- \frac{1}{3} (a_1^2\!+\!a_2^2 \!+\! a_1 a_2) ( a_1 \!+\! a_2 \!-\! f) \ln^3 y \ln (1\!+\!y) 
+ \frac{1}{24} (a_1 \!+\! a_2) (a_1^2 \!+\! a_2^2) \ln^4 y 
\nonumber \\ && \hspace{7mm}
- f (a_1 \!+\! a_2) (3a_1 \!+\! 3a_2 \!-\! 7 f)  \ln (1-y)
\left[ \zeta_3 \!+\! \zeta_2 \ln (1-y) \right]
\nonumber \\ && \hspace{7mm}
+ 2 f (a_1 \!+\! a_2) \zeta_2 \ln (1\!-\!y)
\left[ (a_1 \!+\! a_2 \!-\! 3 f) \ln y 
     + (a_1 \!+\! a_2 \!-\! f)  \ln (1\!+\!y) \right] 
\nonumber \\ && \hspace{7mm}
+  (a_1 \!+\! a_2) (a_1 \!+\! a_2 \!-\! f) \ln (1\!+\!y)
\left[ (4a_1 \!+\! 4a_2 \!-\! 7f)  \zeta_3 + 2 (a_1 \!+\! a_2 \!-\! 3 f) \zeta_2 \ln y \right]
\nonumber \\ && \hspace{7mm}
- \frac{1}{2} (a_1 \!+\! a_2 \!-\! 3f) (a_1^2 \!+\! a_2^2 \!+\! a_1 a_2) \zeta_2 \ln^2 y 
\nonumber \\ && \hspace{7mm}
+ \frac{1}{4} \zeta_4 
\left[ 
45 f^2 (a_1\!+\!a_2) \!-\! 60 f a_1 a_2 \!-\! 9 (a_1^3 \!+\! a_2^3) \!+\! a_1 a_2 (a_1 \!+\! a_2)
\right]
\nonumber \\ && \hspace{7mm}
+ \zeta_3 \ln y 
\left[ 7 f (a_1 a_2 \!-\! f (a_1 \!+\! a_2 ) \!+\! a_1^2 \!+\! a_2^2 )
\!-\! 2 (a_1^3\!+\!a_2^3) \!-\! 3a_1 a_2 (a_1 \!\!+ a_2) \right]
\Biggr\}
\nonumber \\ && \hspace{-3mm}
+ {\cal O} (\ep^4)
\Biggr) \;, 
\label{A_expansion:1}
\end{eqnarray}
and 
\begin{eqnarray}
&& \hspace{-5mm}
\left. _2F_1\left( \begin{array}{c} a_1 \ep, a_2 \ep \\
                 \frac{1}{2} + f \ep \end{array} \right| z\right) 
= 
1 + a_1 a_2 \ep^2 
\Biggl(
- \frac{1}{2} \ln^2 y 
\nonumber \\ && \hspace{-3mm}
+ \ep \Biggl\{
  2 f \left[ 2 \Li{3}{y} - \ln y \Li{2}{y} \right]
+ 2 ( a_1 \!+\! a_2 \!-\! f) \left[ 2 \Li{3}{-y} - \ln y \Li{2}{-y} \right]
\nonumber \\ && \hspace{-3mm}
- \frac{1}{6} (a_1 \!+\! a_2) \ln^3 y 
+ ( a_1 \!+\! a_2 \!-\! 3 f) \zeta_2 \ln y 
+ ( 3 a_1 \!+\! 3 a_2 \!-\! 7 f) \zeta_3
\Biggr\}
\nonumber \\ && \hspace{-3mm}
+ \ep^2 \Biggl\{
4 f (a_1\!+\! a_2 \!-\! 2 f) \ln y  \Snp{1,2}{y} 
\nonumber \\ && \hspace{7mm}
- 2 f (a_1\!+\! a_2 \!-\! f) \ln y  \Snp{1,2}{y^2}
- 4 (a_1\!+\! a_2 \!-\! f) (a_1\!+\! a_2 \!-\! 2 f) \ln y \Snp{1,2}{-y} 
\nonumber \\ && \hspace{7mm}
+ 2 f (a_1 \!+\! a_2) \ln y \Li{3}{y}
+ 2 (a_1\!+\! a_2) (a_1\!+\! a_2 \!-\! f) \ln y \Li{3}{-y} 
\nonumber \\ && \hspace{7mm}
+ 2  \left[ (a_1\!+\! a_2 \!-\! f) \Li{2}{-y} + f \Li{2}{y}\right]^2
\nonumber \\ && \hspace{7mm}
- f (a_1 \!+\! a_2) \ln^2 y \Li{2}{y}
- (a_1 \!+\! a_2) (a_1\!+\! a_2 \!-\! f) \ln^2 y \Li{2}{-y}
\nonumber \\ && \hspace{7mm}
+ 2 (a_1\!+\! a_2 \!-\! 3f) \zeta_2 
\left[ 
f \Li{2}{y} + (a_1\!+\! a_2 \!-\! f) \Li{2}{-y} + \frac{1}{4} (a_1\!+\!a_2)\ln^2 y 
\right]
\nonumber \\ && \hspace{7mm}
- \frac{1}{24} (a_1^2 \!+\! a_2^2 \!+\! a_1 a_2) \ln^4 y 
+ \zeta_3 \ln y \left[ 2 (a_1\!+\!a_2)^2 \!-\! 7 f (a_1\!+\!a_2\!-\!f) \right]
\nonumber \\ && \hspace{7mm}
+ \frac{1}{4} \zeta_4 
\left[ 10 a_1 a_2 \!-\! 30 f (a_1\!+\!a_2) \!+\! 5 (a_1^2 \!+\! a_2^2) \!+\! 45 f^2\right]
\Biggr\}
+ {\cal O} (\ep^3)
\Biggr) \; , 
\label{A_expansion:2}
\end{eqnarray}
where 
\begin{eqnarray}
y = \frac{1-\sqrt{\frac{z}{z-1}}}{1+\sqrt{\frac{z}{z-1}}} \;, \quad 
z = -\frac{(1-y)^2}{4y} \;, \quad
1\!-\! z = \frac{(1+y)^2}{4y} \;, \quad
z \frac{d}{dz} = - \frac{1-y}{1+y} y \frac{d}{dy} \;.
\label{y}
\end{eqnarray}

\subsection{type {\bf B}}
For getting the $\ep$-expansion of a function of {\bf B}-type, 
we will use the representation (\ref{B->A:1}) and (\ref{B->A:2}), 
where r.h.s.'s of the proper equations are given by 
relations (\ref{A_expansion:1}) and (\ref{A_expansion:2}). In this case, the result of 
the $\ep$-expansion can be written in compact form in terms of the variable  $\chi$
related with  variable $y$ (\ref{y}) as following:

\begin{eqnarray}
\chi \equiv \left.y \right|_{z \to 1 \!-\! 1/z} \!=\! \frac{1\!-\!\sqrt{1\!-\!z}}{1\!+\!\sqrt{1\!-\!z}}  
\;, \quad 
z \!=\! \frac{4\chi}{(1+\chi)^2} \;, \quad 
\sqrt{1\!-\!z} \!=\! \frac{1\!-\!\chi}{1\!+\!\chi} \;, \quad
z \frac{d}{dz} \!=\! \frac{1\!+\!\chi}{1\!-\!\chi} \chi \frac{d}{d \chi} \;.
\label{chi}
\end{eqnarray}
The form of the variable  $\chi$ automatically follows from our algebraic relations
and variable $y$ defined by Eq.~(\ref{y}). 
Up to order $O(\ep^3)$ (functions of {\it weight} {\bf 3}) 
the result of $\ep$ expansion of functions of type {\bf B} 
can be cross-checked via {\it multiple binomial sums} studied in \cite{binomial,DK04}.
The results of order $O(\ep^4)$ are new. 

\subsection{type {\bf C}}

In this case we will use the relations (\ref{C->A:1}) and (\ref{C->A:2}) for construction 
of $\ep$-expansion. 
The variable  $y$ is transformed into variable $y_C$ via
\begin{eqnarray}
y_C \equiv \left.y \right|_{z \to -z/(1-z)}  = \frac{1-\sqrt{z}}{1+\sqrt{z}} \; .
\label{y_C}
\end{eqnarray}
The form of a new variable  $y_C$ automatically follows from our algebraic relations
and  the definition of $y$, Eq.~(\ref{y}).
See also discussion in Appendix D of \cite{DK04}.
For this case, the previous results \cite{DK04} allow us to get expansion only 
up to functions of {\it weight} {\bf 2}. The next two orders of $\ep$-expansion are new.

\subsection{type {\bf D}}
For the $\ep$-expansion of {\bf D}-type functions the relations (\ref{D->A:1}) and (\ref{D->A:2}) 
are used. 
In this case we have the original conformal variable (\ref{y}). 
Only logarithmic corrections (functions of {\it weight} {\bf 1}) are available 
from the results of \cite{DK04}.
The next three orders of $\ep$-expansion are new.

\subsection{type {\bf E}}
Functions of type {\bf E} are algebraically independent from previous ones.
Their $\ep$-expansion can be derived by help of relations presented in \cite{FKV}.
%Their $\ep$-expansions are the following (see also Eqs.(D.18) in \cite{DK04}):
It has the following form (see also Eqs.(D.18) in \cite{DK04}):

\begin{eqnarray}
&& 
\left. _2F_1\left( \begin{array}{c} 1+a_1 \ep, 1+a_2\ep \\
                 2 + c \ep \end{array} \right| z\right) 
= 
\frac{1+c\ep}{z}
\Biggl(
- \ln (1-z) 
- \ep \Biggl\{
\frac{c-a_1-a_2}{2} \ln^2 (1-z) + c \Li{2}{z}
      \Biggr\}
\nonumber \\ && 
+ \ep^2 \Biggl\{
  \left[ (a_1+a_2)c - c^2 - 2 a_1 a_2  \right] \Snp{1,2}{z} 
+ \left[ (a_1+a_2)c - c^2 - a_1 a_2    \right]  \ln(1-z) \Li{2}{z} 
\nonumber \\ && \hspace{10mm}
+ c^2 \Li{3}{z}
- \frac{1}{6}(c-a_1-a_2)^2 \ln^3 (1-z) 
      \Biggr\}
\nonumber \\ && 
- \ep^3 \Biggl\{
  c \left[ (a_1+a_2)c - c^2 - 2 a_1 a_2 \right] \Snp{2,2}{z} 
+ c \left[ (a_1+a_2)c - c^2 -   a_1 a_2 \right] \ln(1-z) \Li{3}{z} 
\nonumber \\ && \hspace{10mm}
+ (c-a_1) (c-a_2) (c-a_1-a_2)  \left[  \ln(1-z) \Snp{1,2}{z} + \frac{1}{2}  \ln^2 (1-z) \Li{2}{z} \right]
\nonumber \\ && \hspace{10mm}
+ \frac{1}{24} (c-a_1-a_2)^3 \ln^4 (1-z) 
+ c (c - a_1 - a_2)^2 \Snp{1,3}{z} 
+ c^3 \Li{4}{z} 
\Biggr\}
+ {\cal O} (\ep^4)
\Biggr) \; ,
\label{E-expansion:1}
\end{eqnarray}
\begin{eqnarray}
&& 
\left. _2F_1\left( \begin{array}{c} a_1 \ep, a_2\ep \\
                 1 + c \ep \end{array} \right| z\right) 
= 
1 + a_1 a_2 \ep^2 
\Biggl(
\Li{2}{z}
- \ep \Biggl\{
(c-a_1-a_2) \Snp{1,2}{z} + c \Li{3}{z}
      \Biggr\}
\nonumber \\ && 
+ \ep^2 \Biggl\{
 c^2 \Li{4}{z}
+ (c-a_1-a_2)^2 \Snp{1,3}{z}
+ \frac{1}{2} \left[ c(c-a_1-a_2)+a_1a_2 \right] \left[ \Li{2}{z} \right]^2
\nonumber \\ && \hspace{10mm}
- \left[ c(c-a_1-a_2)+2a_1a_2 \right] \Snp{2,2}{z}
      \Biggr\}
+ {\cal O} (\ep^3)
\Biggr) \; .
\label{E-expansion:2}
\end{eqnarray}
In particular, there are two interesting cases:
%%%%%%%%%%%%%%%%%%%%%%%%%%%%%%%%%%%%%%%%%%%%%%%%%%%%%%%%%%%%%%%
\begin{eqnarray}
&& 
%%%%%%%%%%%%%%%%%%%%%%%%%%%%%%%%%%%%%%%%%%%%%%%%%%%%%%%%%%%%%%%
\left. _2F_1\left( \begin{array}{c} a \ep, b \ep  \\
                 1 + b \ep \end{array} \right| z \right) 
= 1 - \sum_{i=2}^\infty \ep^i \sum_{k=1}^{i-1} a^k (-b)^{i-k} \Snp{i-k,k}{z} \;,
\\ && 
\left. _2F_1\left( \begin{array}{c} 1, -\ep  \\
                 1 - \ep \end{array} \right| z \right) 
= 1 - \sum_{i=1}^\infty \ep^i \Li{i}{z} \;.
\end{eqnarray}
%%%%%%%%%%%%%%%%%%%%%%%%%%%%%%%%%%%%%%%%%%%%%%%%%%%%%%%%%%%%%%%%%%%%%%%%%%%%%%%%%%
\subsection{type {\bf F}}
Let us start from the one-fold integral representation
\begin{eqnarray}
&& \hspace{-10mm} 
\left. _2F_1\left( \begin{array}{c} 
\tfrac{1}{2} \!+\! b_1 \ep, \tfrac{1}{2} \!+\! b_2 \ep  \\
1 \!+\! c \ep  \end{array} \right| z \right) 
= \frac{2 \Gamma(1\!+\!c\ep)}
{\Gamma\left(\tfrac{1}{2}\!+\!b_1 \ep \right) \Gamma\left(\tfrac{1}{2}\!+\!(c\!-\!b_1) \ep \right)}
\int_0^{\pi/2} d \theta 
\frac{(\sin \theta)^{2b_1\ep} (\cos \theta)^{2(c\!-\!b_1)\ep}}
{(1\!-\!z\sin^2 \theta)^{1/2+b_2\ep}} \;,
\nonumber \\ && \hspace{-10mm} 
\left. _2F_1\left( \begin{array}{c} 
\tfrac{3}{2} \!+\! b_1 \ep, \tfrac{3}{2} \!+\! b_2 \ep  \\
2 \!+\! c \ep  \end{array} \right| z \right) 
= \frac{2 \Gamma(2\!+\!c\ep)}
{\Gamma\left(\tfrac{3}{2}\!+\!b_1 \ep \right) \Gamma\left(\tfrac{1}{2}\!+\!(c\!-\!b_1) \ep \right)}
\int_0^{\pi/2} d \theta 
\frac{(\sin \theta)^{2\!+\!2b_1\ep} (\cos \theta)^{2(c\!-\!b_1)\ep}}
{(1\!-\!z\sin^2 \theta)^{3/2+b_2\ep}} \;.
\nonumber \\ && \hspace{-10mm} 
\label{f}
\end{eqnarray}
The finite part of the first integral is equal to complete elliptic integral of the first kind $K(k)$ 
defined as 
\begin{eqnarray}
K(k) & \equiv & F\left(\frac{\pi}{2},k \right) 
= \int_0^{\pi/2} \frac{d \phi}{\sqrt{1-k^2 \sin^2 \phi}} 
= \frac{\pi}{2}
\left. _2F_1\left( \begin{array}{c} 
\tfrac{1}{2}, \tfrac{1}{2}  \\
1  \end{array} \right| k^2 \right) \;.
\end{eqnarray}
The finite part of the second hypergeometric function is expressible in terms of 
the complete elliptic integrals of the first and second kind \cite{ww}. Using the relation 
\begin{eqnarray}
&& \hspace{-10mm}
\frac{( 1 \!+\! 2b_1 \ep) ( 1 \!+\! 2b_2 \ep)}{2 (1\!+\!c\ep)}
z (1\!-\!z)
\left. _2F_1\left( \begin{array}{c} 
\tfrac{3}{2}\!+\!b_1\ep, \tfrac{3}{2}\!+\!b_2\ep  \\
2\!+\!c\ep  \end{array} \right| z \right) 
\nonumber \\ && 
= 
\left[1+2(c-b_1) \ep \right]
\left. _2F_1\left( \begin{array}{c} 
-\tfrac{1}{2}\!+\!b_1 \ep, \tfrac{1}{2} \!+\! b_2 \ep  \\
1\!+\!c\ep  \end{array} \right| z \right) 
\nonumber \\ && 
+
\left[ ( 1 \!+\! 2b_2 \ep)z - (1 + 2 (c-b_1) \ep) \right]
\left. _2F_1\left( \begin{array}{c} 
\tfrac{1}{2}\!+\!b_1 \ep, \tfrac{1}{2} \!+\! b_2 \ep  \\
1\!+\!c\ep  \end{array} \right| z \right) \;,
\end{eqnarray}
and definition of the complete elliptic integral of the second kind $E(k)$,
\begin{eqnarray}
E(k) & \equiv & E\left(\frac{\pi}{2},k \right) 
= \int_0^{\pi/2} d \phi \sqrt{1-k^2 \sin^2 \phi} 
= 
\frac{\pi}{2}
\left. _2F_1\left( \begin{array}{c} 
-\tfrac{1}{2}, \tfrac{1}{2}  \\
1  \end{array} \right| k^2 \right) \;,
\end{eqnarray}
we get 
\begin{eqnarray}
\left. _2F_1\left( \begin{array}{c} 
\tfrac{3}{2}, \tfrac{3}{2}  \\
2  \end{array} \right| z \right) 
= 
\frac{4}{\pi z(1\!-\!z)}
\left[ E(\sqrt{z}) - (1\!-\!z) K(\sqrt{z}) \right] \;.
\end{eqnarray}
The next coefficients in the $\ep$-expansion of the functions (\ref{f}) are related to 
some generalization of elliptic functions \cite{generalization}.
In terms of the multiple sums, these new functions are related to 
{\it multiple double sums} defined as 
\begin{equation}
\sum_{j=1}^\infty \left( 2j \atop j\right)^2  \frac{u^j}{j^c}
S_{a_1} \ldots S_{a_p} \bar{S}_{b_1} \ldots \bar{S}_{b_q} \; , 
\label{double}
\end{equation}
In the rest of present paper, we omit functions of type {\bf F} from our consideration.
For the definition of coefficients of the $\ep$-expansion of functions of type {\bf F}, 
the one-fold integral representation 
the (\ref{f}) or representation of type (\ref{integral_representation}) and/or (\ref{integral_representationII})
should be used.

%%%%%%%%%%%%%%%%%%%%%%%%%%%%%%%%%%%%%%%%%%%%%%%%%%%%%%%%%%%%%%%%%%%%%%%%%%%%%%%%%%
\section{Application to Feynman diagrams}
There are several important master-integrals expressible in terms of 
${}_2F_1$ hypergeometric functions.
This set of integrals includes 
one-loop propagator type diagram with arbitrary values of mass and momentum \cite{DK01,D-ep};
two-loop bubble integral with an arbitrary values of masses \cite{DK01,D-ep}, and 
one-loop massless vertex diagram with three non-zero external momenta \cite{DK01}.
For these diagrams, all order $\ep$-expansions can be written in terms of Nielsen polylogarithms only \cite{DK01}.
Our technology for the expansion of hypergeometric functions has been applied also in 
\cite{binomial,DK04,qcd} to more complicated case of generalized hypergeometric function, 
like ${}_3F_2$ and  ${}_4F_3$.
The two-loop propagator type diagram ${\bf V1001}$ and three-loop bubble-type diagram ${\bf E_{3}}$
(see \cite{DK01} for details) are expressible in terms of  ${}_2F_1$-functions of argument 1/4. 
In this case, the result of expansion can be written in terms 
of generalized log-sine functions and calculated with high accuracy by the help 
of program {\bf LSJK} \cite{lsjk}. 

\begin{figure}[th]
\begin{center}
\centerline{\vbox{\epsfysize=45mm \epsfbox{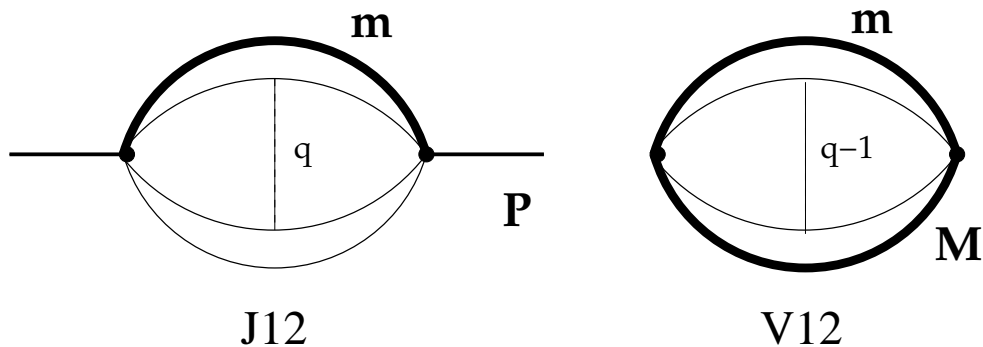}}}
\caption{\label{diagrams} Diagrams considered in the paper.
Bold and thin lines correspond to massive and
massless propagators, respectively.}
\end{center}
\end{figure}

Below we present some diagrams where the results are expressible in 
terms of Gauss hypergeometric functions for an arbitrary set of indices.
The proper diagrams are shown in Fig.~\ref{diagrams}.
Diagrams of this type suffer, in general, from irreducible numerator, 
so that the solution of recurrence relations is 
nontrivial problem (besides one-loop propagator and two-loop bubble cases).
The solution of recurrence relations for two-loop sunset-type diagram 
was presented in \cite{FJTV,programs:propagator}. 
Using the algorithm \cite{davydychev,tarasov}, 
any tensor integral can be presented in terms of scalar integrals 
with the shifted space-time dimension and arbitrary (positive) powers of propagators.
In scalar integrals of given type, massless subloops can be integrated, and the original 
integrals effectively reduce to more simple integrals with some powers of propagators 
shifted by terms proportional to $\ep$. However, for the gauge invariance 
reason, it is desirable to reduce all diagrams to a set of master-integrals before 
construction of $\ep$-expansion.

Using the reduction algorithm described in Sec.~2 it is possible to 
express the arbitrary tensor integral of given type in terms of our basis 
functions (master-integrals) and integrals of more simple structure.
The $\ep$-expansion of master-integrals can be done by help of  relations presented in Sec.~4.
We would like to note that we are working in Euclidean space-time $(p^2=-M^2)$
and use the normalization that each loop is divided by a factor $\Gamma(1+\ep)$.

%%%%%%%%%%%%%%%%%%%%%%%%%%%%%%%%%%%%%%%%%%%%%%%%%%%%%%%%%%%%%%%%%%%
\subsection{$q$--loop propagator with $q$ massless line}
Let us consider the $q$-loop sunset-type propagator with one massive line 
and $q$ massless ones. The result is expressible in terms of 
Gauss hypergeometric function 
\begin{eqnarray}
&&   \hspace{-4mm}
J_{12}(\alpha_1, \alpha_2, \cdots, \alpha_q, \beta, m^2, p^2) = 
\nonumber \\ && \hspace{-4mm}
\frac{1}{\Gamma^q \left( 3 \!-\! \frac{n}{2}\right) }
\int \frac{d^n k_1 d^n k_2 \cdots d^n k_q}{
          [k_1^2+m^2]^{\beta} [(k_1-k_2-\cdots -k_q-p)^2]^{\alpha_1} [k_2^2]^{\alpha_2} \cdots  [k_q^2]^{\alpha_q}}
\nonumber \\ && \hspace{-4mm}
= 
\left[ \Pi_{l=1}^q \frac{\Gamma \left( \frac{n}{2}-\alpha_l \right)}{\Gamma(\alpha_l)} \right]
%(-1)^{\beta+\alpha} 
\pi^{qn/2}
(m^2)^{qn/2 \!-\! \beta \!-\!\alpha}
\frac{\Gamma \left( \alpha \!+\! \beta \!-\! \frac{n}{2}q \right) 
      \Gamma \left( \alpha \!-\! \frac{n}{2}(q-1) \right)}
     {\Gamma(\beta) \Gamma \left( \frac{n}{2} \right) \Gamma^q \left( 3 \!-\! \frac{n}{2}\right) }
\nonumber \\ && 
\times
{}_{2}F_1\left(\begin{array}{c|}
\alpha - \tfrac{n}{2}(q-1), \alpha \!+\! \beta \!-\! \tfrac{n}{2}q \\
\tfrac{n}{2} \end{array} ~ - \frac{p^2}{m^2} \right) \;,
\end{eqnarray}
where 
$$
\alpha = \sum_{r=1}^q \alpha_r \;.
$$
For given type of diagram there are only two nontrivial master-integrals.
In the parametrization $n=2m-2\ep$ with integer $m$, the basis is 
\begin{eqnarray}
{}_{2}F_1\left(\begin{array}{c|}
1 \!+\! \ep (q-1), 1 \!+\! \ep q\\
2-\ep \end{array} ~ - \frac{p^2}{m^2} \right) \;, 
\quad 
{}_{2}F_1\left(\begin{array}{c|}
\ep (q-1), \ep q \\
1-\ep \end{array} ~ - \frac{p^2}{m^2} \right) \;.
\end{eqnarray}
The two-loop case $(q=2)$ has been considered in \cite{FJTV}.
%
%
%%%%%%%%%%%%%%%%%%%%%%%%%%%%%%%%%%%%%%%%%%%%%%%%%%%%%%%%%%%%%%%%%%%
\subsection{$q$--loop bubble with $q-1$ massless lines}
Let us consider now the $q$-loop bubble with two massive lines and $q-1$ massless lines. The result is 
\begin{eqnarray}
&&   \hspace{-4mm}
V_{12}(\alpha_1, \alpha_2, \cdots, \alpha_p, \beta_1, \beta_2, m^2, M^2) = 
\nonumber \\ && \hspace{-4mm}
\frac{1}{\Gamma^q \left( 3 \!-\! \frac{n}{2}\right) }
\int \frac{d^n k_1 d^n k_2 \cdots d^n k_q}{
          [k_p^2+m^2]^{\beta_1} [(k_{p-1}^2+M^2]^{\beta_2} [k_1^2]^{\alpha_1} \cdots  [k_{p-2}^2]^{\alpha_{p-2}} 
          [(k_p-k_1-k_2-\cdots-k_{p-1})^2]^{\alpha_{q-1}} }
\nonumber \\ && \hspace{-4mm}
= 
\left[ \Pi_{l=1}^{q-1} \frac{\Gamma \left( \frac{n}{2}-\alpha_l \right)}{\Gamma(\alpha_l)} \right]
\frac{\pi^{qn/2}(m^2)^{qn/2 \!-\! \beta_1 \!-\! \beta_2 \!-\!\alpha} }
     {\Gamma(\beta_1) \Gamma(\beta_2) \Gamma \left( \frac{n}{2} \right) \Gamma^q \left( 3 \!-\! \frac{n}{2}\right) }
\nonumber \\ && 
\times
\Biggl\{
\Gamma \left( \frac{n}{2} \!-\! \beta_2 \right)
\Gamma \left( \alpha \!+\! \beta_2 \!-\! \frac{n}{2}(q-1) \right)
\Gamma \left( \alpha \!+\! \beta_1 \!+\! \beta_2 \!-\! \frac{n}{2} q \right)
\nonumber \\ && \hspace{4mm}
\times
{}_{2}F_1\left(\begin{array}{c|}
\beta_1 \!+\! \beta_2 \!+\! \alpha \!-\! \tfrac{n}{2}q, \alpha \!+\! \beta_2 \!-\! \tfrac{n}{2} (q-1) \\
1 \!+\! \beta_2 \!-\! \tfrac{n}{2} \end{array} ~ \frac{M^2}{m^2} \right) 
\nonumber \\ && 
+ 
\Gamma \left( \beta_2 \!-\!\frac{n}{2} \right)
\Gamma \left( \alpha \!+\! \beta_1 \!-\! \frac{n}{2}(q-1) \right)
\Gamma \left( \alpha \!-\! \frac{n}{2}(q-2) \right)
\left(\frac{M^2}{m^2} \right)^{n/2-\beta_2}
\nonumber \\ && \hspace{4mm}
\times
{}_{2}F_1\left(\begin{array}{c|}
\alpha \!-\! \tfrac{n}{2}(q-2), \alpha \!+\! \beta_1 \!-\! \tfrac{n}{2}(q-1) \\
1 \!-\! \beta_2 \!+\! \tfrac{n}{2} \end{array} ~ \frac{M^2}{m^2} \right) 
\Biggr \} \;,
\end{eqnarray}
where 
$$
\alpha = \sum_{r=1}^{q-1} \alpha_r \;.
$$
The results of the reduction are expressible in terms of 
four Gauss hypergeometric functions. In the parametrization $n=2m-2\ep$, 
where is $m$ is an integer number we get four basis functions:
\begin{eqnarray}
&& 
{}_{2}F_1\left(\begin{array}{c|}
1 \!+\! \ep (q\!-\!1), 1 \!+\! \ep q\\
2+\ep \end{array} ~ \frac{M^2}{m^2} \right) \;, 
\quad 
{}_{2}F_1\left(\begin{array}{c|}
\ep (q\!-\!1), \ep q \\
1+\ep \end{array} ~ \frac{M^2}{m^2} \right) \;,
\nonumber \\ && 
{}_{2}F_1\left(\begin{array}{c|}
1 \!+\! \ep (q\!-\!2), 1 \!+\! \ep (q\!-\!1) \\
2-\ep \end{array} ~  \frac{M^2}{m^2} \right) \;, 
\quad 
{}_{2}F_1\left(\begin{array}{c|}
\ep (q\!-\!2), \ep (q\!-\!1) \\
1-\ep \end{array} ~  \frac{M^2}{m^2} \right) \;.
\end{eqnarray}
Only for $q=2$ (two-loop case) 
these four hypergeometric functions are expressible in terms 
of one Gauss hypergeometric function and the function ${}_1F_0$, so that only one 
nontrivial master-integral exists. It was calculated in \cite{v012}.
For $q>2$ (3-loop or more) there are four independent Gauss hypergeometric functions.
As a consequence, there are four nontrivial master-integrals for diagrams of this type 
at 3-loop or more.
%%%%%%%%%%%%%%%%%%%%%%%%%%%%%%%%%%%%%%%%%%%%%%%%%%%%%%%%%%%%%%%%%%%%%%%%%%%%%%%%%%
\section{Conclusion}

In this paper we have presented the reduction algorithm for 
Gauss hypergeometric functions with arbitrary values of parameters 
to the two functions (\ref{basis-function}) with fixed values of parameters,
which differ from original ones by integers.

It was shown that the Gauss hypergeometric functions with integer/half-integer 
values of parameters can be divided into 6 types (see Table~I). 
Only three type of them are  algebraically independent. 
We have presented the explicit relations which allow us to express the 
functions of type {\bf B, C, D} in terms of functions 
of type {\bf A} (see Eqs.~(\ref{D->A:1}), (\ref{D->A:2}), (\ref{C->A:1}), 
(\ref{C->A:2}), (\ref{B->A:1}), (\ref{B->A:2}) ).
For functions of type {\bf A, B, C, D, E} the higher-order $\ep$-expansion 
up to functions of {\it weight} {\bf 4} are constructed (see Eqs.
(\ref{A_expansion:1}), (\ref{A_expansion:2}), (\ref{E-expansion:1}),(\ref{E-expansion:2}) ).
The result of the expansion is expressible in terms of Nielsen polylogarithms only. 
The $\ep$-expansion of function of type {\bf F} is expressible in terms 
of new functions related to generalizations of elliptic functions. 

As an illustration of the application of our algorithm of reduction, we have considered the 
reduction of $q$--loop off-shell propagator diagrams with one massive line and 
$q$ massless lines and $q$--loop bubble with two-massive lines and $q-1$ massless lines.
We demonstrated, that the number of master-integrals for sunset-type diagram beyond one-loop 
does not depend on the number of internal massless lines and it is equal to two.
For bubble type-diagram beyond two-loop, the number of master-integrals is equal to four.
%%%%%%%%%%%%%%%%%%%%%%%%%%%%%%%%%%%%%%%%%%%%%%%%%%%%%%%%%%%%%%%%%%%%%%%%%%%%%%%%%%

{\bf Acknowledgements.}
I would like to thank A.~Sheplyakov for checks of some formulae. 
I am grateful A.~Davydychev for useful discussion
and for careful reading of the manuscript.
Many thanks to S.~Yost for reviewing the manuscript.
I am thankful to E.~Cardi and M.~Rogal for correspondence about some typos in earlier versions. 

%%%%%%%%%%%%%%%%%%%%%%%%%%%%%%%%%%%%%%%%%%%%%%%%%%%%%%%%%%%%%%%%%%%%%%%%%%%%%%%%%%
%%%%%%%%%%%%%%%%%%%%%%%%%%%%%%%%%%%%%%%%%%%%%%%%%%%%%%%%%%%%%%%%%%%%%%%%%%%%%%%%%%

%%%%%%%%%%%%%%%%%%%%%%%%%%%%%%%%%%%%%%%%%%%%%%%%%%%%%%%%%%%%%%%%%%%%%%%%%%%%%%%%%%%%%%%%%
\end{document}